# Explaining Cognitive Computing Through the Information Systems Lens

*Completed Research Paper*

**Samaa Elnagar**
Virginia Commonwealth University
Richmond, Virginia, USA
elnagarsa@vcu.edu

**Manoj A. Thomas**
University of Sydney
Sydney, Australia
manoj.thomas@sydney.edu.au

## Abstract

Cognitive computing (COC) aims to embed human cognition into computerized models. However, there is no scientific classification that delineates the nature of Cognitive Computing. Unlike the medical and computer science fields, Information Systems (IS) has conducted very little research on COC. Although the potential to make important research contributions in this area is great, we argue that the lack of a cohesive interpretation of what constitutes COC has led to inferior COC research in IS. Therefore, we need first to clearly identify COC as a phenomenon to be able to identify and guide prospective research areas in IS. In this research, a phenomenological approach is adopted using thematic analysis to the published literature in COC research. Then, we discuss how IS may contribute to the development of design science artifacts under the COC umbrella. In addition, the paper raises important questions for future research by highlighting how IS researchers could make meaningful contributions to this emerging topic.

**Keywords:** Cognitive Computing, Neuromorphic Engineering, Artificial Intelligence, Information systems research, Thematic analysis, Phenomenology.

## Introduction

The ultimate goal of Cognitive Computing (COC) is to empower machines with human intelligence (Wang et al. 2009). The first era of computing started in the early 1900s and was called the *Tabulating Era* where punch cards were used to instruct the machine to perform calculations. Then followed the *Programming Era* in the 1950s (Guan et al. 2007). COC is considered the third era of computing (Hildesheim 2018).

Brynjolfsson and McAfee (2012 highlighted that organizations need cognitive systems that can handle complexity, make confidence-based predictions, learn actively and passively, act autonomously, and reflect a well-scoped purpose. On the other hand, some organizations expect that COC could cut costs and increase productivity by automating many manual intensive jobs (Hildesheim 2018). COC is not only cognitive software applications but also low-power behaving hardware. COC hardware extends traditionally computerized systems to brain-like machines (Indiveri et al. 2009). Cognitive robotics are developed through universal brains that can surpass the human brain performance in a wide variety of tasks (Browne 1997; Wang et al. 2020).

COC based systems are called "next generation AI" because the AI components of COC are no longer a "black box" (IBM 2019). Instead, AI in COC allows more transparent learning by applying predefined intelligence rules in mathematical format (Semantic-Web 2019). COC-based systems could radically change the organizational and operational configurations. However, IS research targeting the COC paradigm is far behind the research conducted in other scientific disciplines such as medicine and engineering. For example: the electrical engineering science is producing NE circuits that mimic the brain processing system. Computer science is developing new Neuromorphic machine learning (NML) techniques. Psychology and applied mathematics are developing the symbolic logic of cognitive activities (Lu and Li 2020).

By the time that giant companies such as Google, IBM, Microsoft are building high-tech cognitive systems for many organizations, ISR still considers COC as the future. Many organizations are adopting COC in their core processes such as Tesla, Progressive, and Deloitte (Radar 2019). In other words, IS research outdated the changes brought to the organizational and economic structures as a result of COC. Therefore, a clear architecture for COC would help IS start conducting meaningful research in COC.

This research is motivated by the lack of unified interpretation of what constitutes COC. Literature shows different discourses around COC. Scholarship that focuses on Artificial Intelligence (AI), mathematics, and brain-like electronics belonging to COC research (Lu and Li 2020). In addition, the lack of IS research in COC is directly driven by the indistinguishable COC determinations. This research provides an explanatory systematic literature review that delineates the nature of COC research. In early attempts to build a welldefined foundation that unify the conflicting understandings of COC.

This paper aims to explore and scrutinize the structure of COC as an emergent phenomenon to be explored, and recommends a rationalization for developing COC research foundations in IS. The paper elaborates COC research areas and the purpose it may serve to the IS research community. The study recommends foreseeable research directions for the IS discipline. The research significance is to comprehensively integrate the dispersed research in COC and to provide guidance for researchers especially Design Science (DS) researchers to undertake a purposeful research agenda in COC.

## Methodology

The objective of this research is to collect evidence about the structure and nature of COC since there is no coalesced characterization of COC. The COC structure is considered the guidance for IS to conduct thoughtful research in COC. We followed a systematic literature review methodology along with thematic analytics to develop the structure of COC. We combined the methods developed by Vesey and Wolfswinkel (Vessey et al. 2002; Wolfswinkel et al. 2013), and shadow the following four steps: (1) identify the scope of the study, (2) search for relevant articles, (3) analyze the literature using qualitative thematic analysis, and (4) synthesize findings to determine the COC structure.

The literature review methodology is shown in Figure 1. Starting from the left, the four main steps are sequentially followed. Then, for each step, essential processes are performed. For example, in defining the scope, research goals and literature review methodology is defined. In each process, specific tools are used. For example, NVivo 12 and EndNote software were used in the qualitative analysis. Moreover, each process produces different outputs that are combined together to build the structure of COC.

### *Scope Identification*

Defining the scope is the first stage in the systematic literature review. It defines what to include or exclude from the study (Jones and Gatrell 2014). The scope or the goal of this literature review is to inductively investigate the COC structure as the first step towards grounded IS research in COC. During the search process we want to ensure "systematicity" through comprehensively documenting the search process (Rowe 2014). Systematicity ensures replicability and enables finding interdependencies so it is easy to extend and criticize (Brust et al. 2017).

### *Search Process*

Since there are diverse disciplines conducting research in COC, building a cohesive view of COC requires integrating literature from all related disciplinary areas. Our search process had three rounds. In the first round, we defined the keywords that form the foundational basis of COC. For our search, we included the following keywords: "Cognitive Computing", "Cognitive Computers", "Intelligence Augmentation", and "Cognitive machine Learning", "Cognitive Science", and "Intelligence Science".

The scientific search databases used included Google Scholar, AIS library, and ProQuest. The initial research round resulted in 137 papers dated from 1990s to 2019. However, 13 papers were excluded which were merely Psychology or Neuroscience oriented and not related to technology or IS.

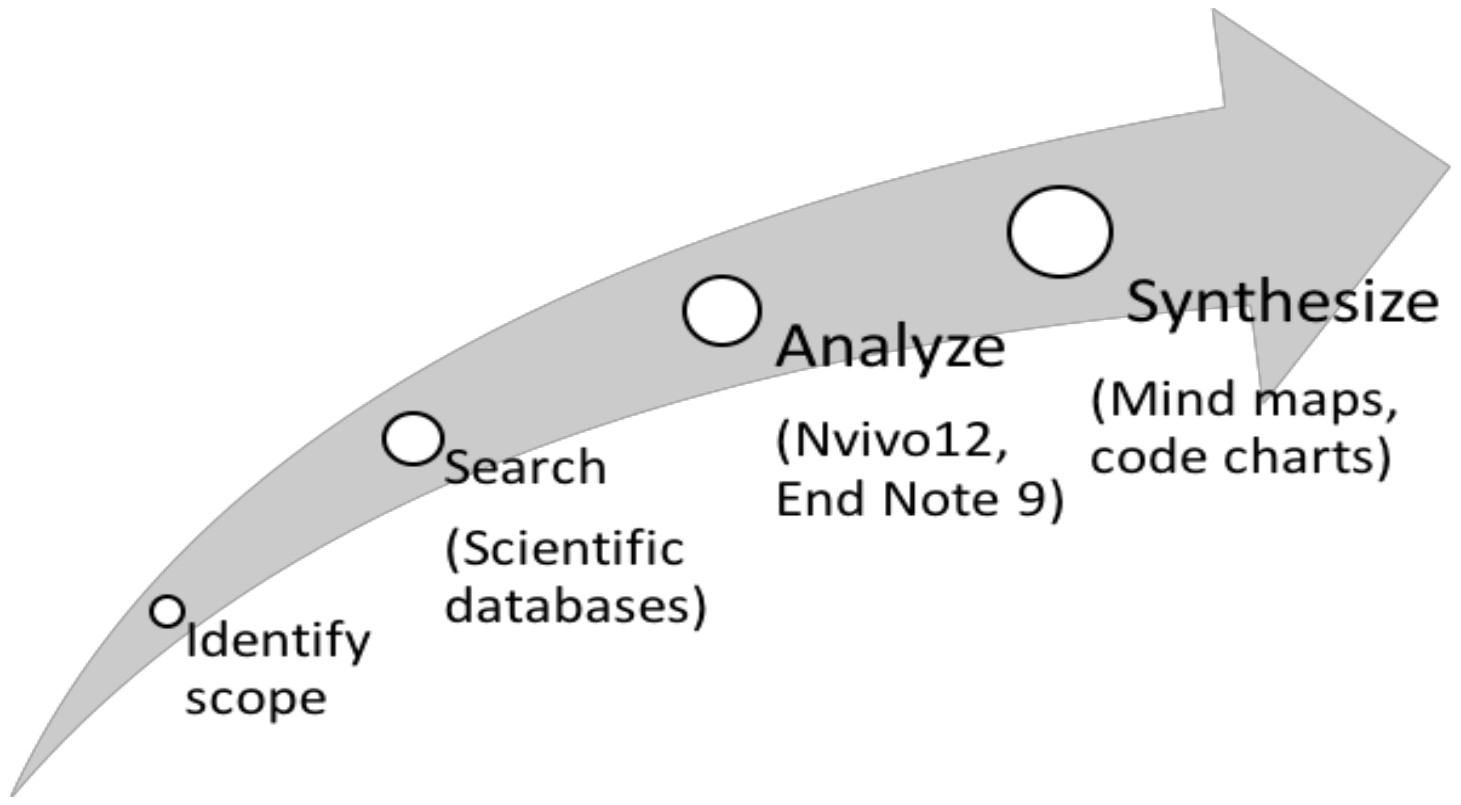

**Figure1. Systematic Literature Review Methodology.**

Due to the novelty of the topic, we did not expect to find mature praxis of COC in IS journals and conferences. However, our literature review process included searching in salient IS journals and databases to ensure a comprehensive search process. A second extended research round was performed that aimed to investigate the implementation of cognitive systems using COC. The keywords used in this round included - "*Cognitive Information Systems*", *"Cognitive Systems"*. The research included but not limited to IS journals. Following Webster and Watson (2002, we conducted special research using the keywords in the basket of eight journal and top AIS conferences.

The second round resulted in 45 additional papers. However, many papers that address cognitive systems are not related to COC. So, Papers that did not include the term "*Cognitive Computing*" as a design principle were excluded. In addition, systems with older and outdated technologies and approaches such as expert systems and ordinary neural networks-based systems were also excluded. This resulted in 27 papers.

We then performed a quick categorization of papers and we found that research in logic related papers such as mathematics, informatics, and history of intelligence contributed 34% of the research papers while hardware related research contributed only 18 % of research papers. Nevertheless, the lower percentage of hardware related papers is deceiving and not to be interpreted as the lack of research in brain-like hardware. The likely reason that hardware represents only 18% of the papers is that recent studies on the development of new brain-like circuits did not consider COC as a related keyword term or research objective. Therefore, we decided to include brain-like hardware related research to our corpus. So, an additional third research round was conducted that aims at investigating brain-like hardware using the terms "*Neuromorphic Engineering*", and "*Neuromorphic neural networks*". However, papers that are of exploratory research and intervention research types were excluded because our focus was not on the technicality of *Neuromorphic circuits*. Rather, we need established research that contributes to COC research. This search round resulted in 47 additional papers.

In the fourth round we conducted backward search and forward search (Jones and Gatrell 2014). In backward search, we inspected where the author cited seminal references to COC, and we correspondingly added these references to the corpus. In forward search, we looked for studies which cited the seminal references using "cited by" feature of Google Scholar and the Web of Science. This ensured the currency of our literature review process. The fourth round added 35 additional papers as summarized in Figure 2. In total, the search process resulted in 227 articles and books retrieved from 56 journals, 55 conferences and workshops. Details about paper references, publishers are found in the Appendix**.**

We applied rejection criteria in each round of the search process as discussed above. Our criteria of inclusion were to include any study that enlightens the foundation of COC including research about logic/ mathematics of cognitive activities, cognitive informatics, brain informatics, cognitive systems, brain-like hardware, and AI algorithms. We saved the meta-knowledge of resulting files (year, publisher, keywords, and author) along with the main focus of the paper (hardware, software, logic, or algorithms) to an excel file for ease of conducting analysis. Then, we created a compilation of references in an EndNote library.

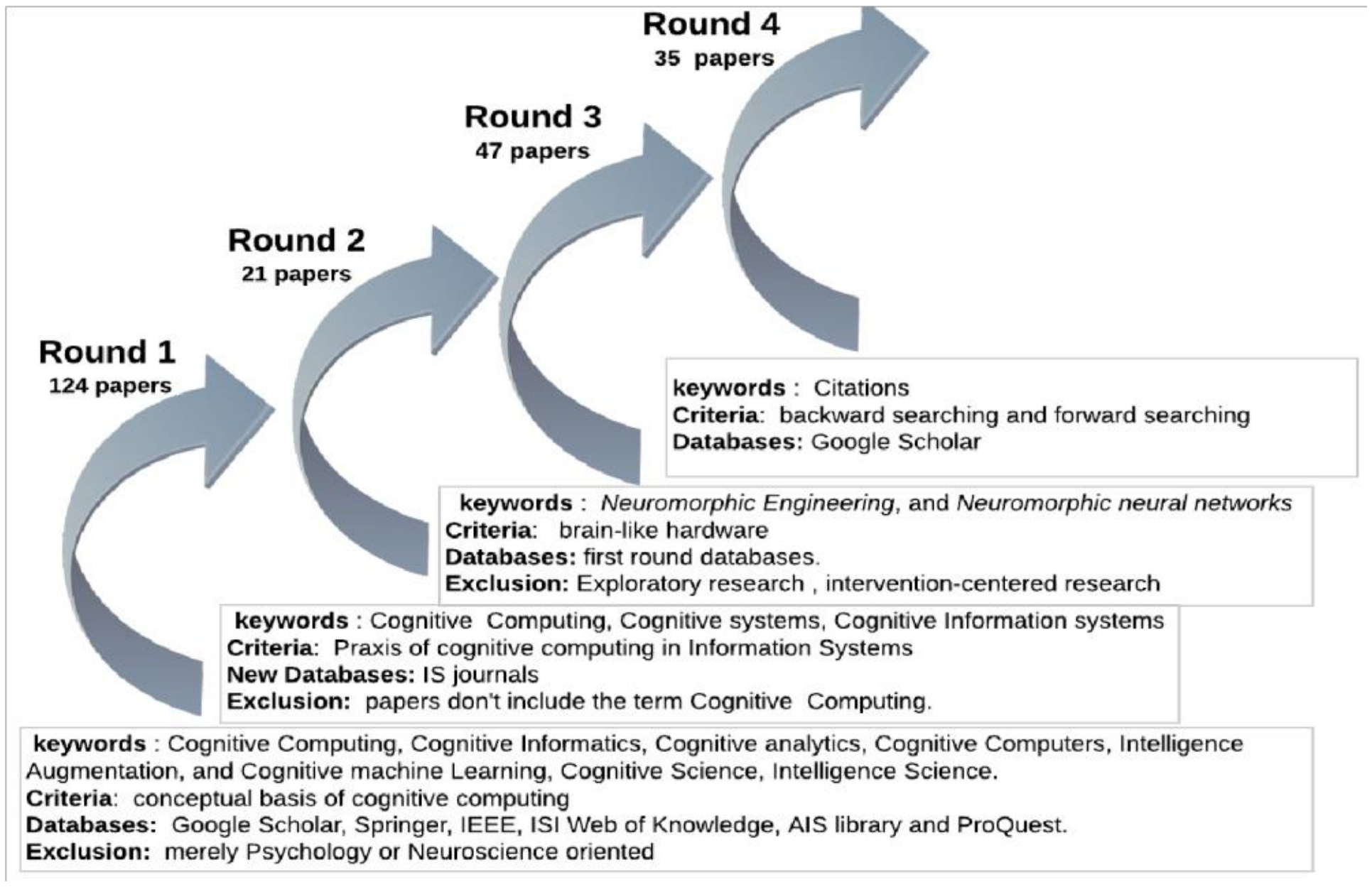

**Figure 2. The searching and selection process.**

## *Analysis*

The analysis phase should be described in detail to allow other researchers to critically evaluate the validity of the authors' inferences (Moher et al. 2009). We used thematic analysis to analyze the corpus and developed the basic building blocks of the COC structure. Thematic analysis is a flexible analysis methodology. Thematic analysis can be used to analyze most types of qualitative data such as those assimilated through action research, interviews, and surveys with unlimited size restrictions.

So why did we choose thematic analysis? Firstly, the nature of the research problem itself. Research about the nature and structure of COC can be considered a *phenomenological* research where research is oriented towards discovery rather than predefined assumptions or deductive goals to be established. *Phenomenology* escapes from judgment on the foundations of absolute truth or falsehood. Rather, *Phenomenology* explains the characteristic consistency in a phenomena (Wisse 2003). In addition, the *malleability* of thematic analysis makes it a more upfront choice than approaches informed by theoretical assumptions (Aronson 1995).

## *Thematic Analysis*

Thematic analysis (TA) aims at finding common patterns (or "themes") within qualitative data. The interpretive power of TA lies in exploring explicit and implicit meanings within the data (Braun and Clarke 2012).

The most popular TA approach is coding by tagging elements of interest with a coding label. The TA process begins with theme development, where the elements of interest are allocated to predefined themes. Our goal is to inductively define the COC structure. So, the codes should emerge through the TA process. Then, the codes themselves would describe the objects to be studied (Coffey and Atkinson 1996).

When applying TA as an inductive approach, the themes identified during the process of coding should comprehensively describe the data without trying to fit the data into a pre-existing theory or framework. The TA process in our inductive approach consists of four phases: data reduction, generating initial themes, grouping codes, and reporting findings (Braun and Clarke 2013). The analysis tool used to extract themes from research papers was Nvivo 12 qualitative research software. The Nvivo codes are found in Appendix[1]

*Data Reduction*: is where data is reduced to themes or categories to ease the identification of data segments that share a common theme or code. Data reduction is dependent on noticing relevant text about phenomena then analyzing the text to find patterns and overlying structures (Coffey and Atkinson 1996).

*Generating Initial Themes*: combining overarching themes in the data. At the beginning, coding was conducted to find main concepts in COC. Nvivo coding resulted in 89 codes that were mapped into a mind map as shown in Figure 3.

*Grouping Codes*: where codes are further grouped into parent codes that represent the top-level class. The 89 codes were stored in the form of nodes and grouped into 15 parent nodes. The nodes and their frequencies are listed in Table 1. As shown in the table, 72 % of the coded nodes referred to cognitive computing definitions, Role, nature. However, the highly frequent parent nodes were: *Logic* or encoding cognitive functions as logical functions (46%), *AI* including deep learning and spike neural networks had (48% ) frequency, Neuromorphic Engineering that aims at developing brain-like hardware (36%), and finally cognitive systems such as IBM and Siri with (40%)

*Reporting Findings:* The 15 parent nodes were further clustered to abstract level to delineate research categories in COC that results in four clusters as follows: the first is research on the logics of intelligence or the mathematical symbolic representations encompassing algorithms, logics and theories such as the work in (Wang 2008) (Ogiela and Ogiela 2018) (Wang 2012). The second line of research is mimicking the physical structure of the brain in the form of electronic circuits or what is called *Neuromorphic circuits* (Mead 1990) (Pfeil et al. 2013). It is to be noted that this line of research is different from the logics and the structure of traditional Von Neumann computing (Indiveri et al. 2009).

The third and most salient research line in COC is AI algorithms, especially deep learning as they offer revolutionary cognitive inference capabilities for a variety of cognitive tasks related to speech and vision (Elnagar and Thomas ; Szegedy et al. 2015). The fourth line of research in COC is cognitive systems (such as the work in (Dessì et al. 2019) (Paletta et al. 2018)), and associated with systems that adapt and evolve automatically in different domains.

1 Literature Review Appendix

| **Codes** | **Themes** | **Frequency** | **No. Child Nodes** |
|---|---|---|---|
| Algorithm | Alg | 9% | 0 |
| Artificial Intelligence | AI | 48% | 9 |
| Cognitive Computers (*CogCs*) | CogCs | 2% | 0 |
| Cognitive Computing | COC | 72% | 10 |
| Cognitive information systems | CIS | 1% | 0 |
| Cognitive systems | COCS | 40% | 6 |
| Information fusion | IF | 5% | 3 |
| Intelligence Augmentation | IA | 4% | 0 |
| Logic | LO | 46% | 15 |
| Mathematics | MATH | 15% | 3 |
| Neuromorphic Engineering | NE | 36% | 17 |
| Programming paradigm | PR | 6% | 3 |
| Sciences (neuro, psychology..etc) | SCs | 17% | 5 |
| Technologies | Tech | 8% | 9 |
| Theories | Th | 8% | 0 |

**Table 1. Concepts Frequency in Cognitive Computing Research.**

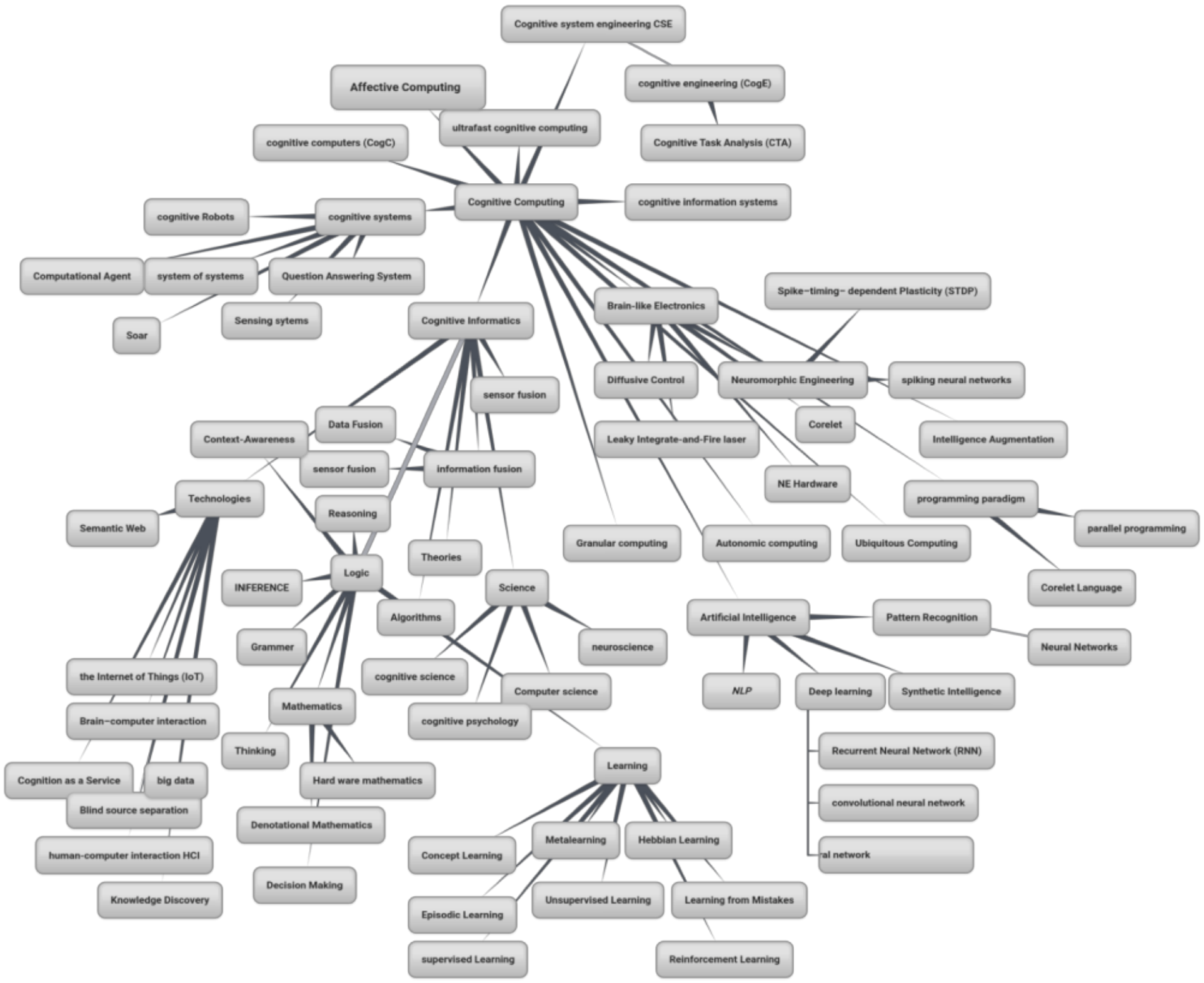

**Figure 3. Mind Map of Nodes Coded based on Thematic Analysis**

## *Synthesize Findings*

Synthesizing the findings from the analysis, we can conclude that research in COC as a phenomenon is diverse and includes research lines on logic of intelligence, hardware that mimics the brain, AI algorithms, and development of cognitive systems. Research on logic of intelligence in the form of *Denotational Mathematics (DM)* was used to describe mental and behavioral activities such as *Visual Semantic Algebra (VSA)* and *Inference Algebra* (Wang 2012). Another line of research aims to produce brain-like hardware or *Neuromorphic circuits* that use spikes – low discrete events, the size of a few bits that convey very little besides the specific time of the spike – as a form of communication. *Neuromorphic circuits also* collocate memory with processing instead of separate storage units in traditional computers (Nahmias et al. 2013).

AI algorithms have received notable attention in COC research, especially deep learning algorithms such as convolutional networks and recurrent networks (Mordvintsev et al. 2015). Interestingly, research on AI algorithms tends to target both *Neuromorphic circuits* and Von Neumann circuits. There are also the recurrent deep networks for *Neuromorphic circuits* called *Spiking Recurrent Networks* (Cao et al. 2015) that are designed to run on neuromorphic circuits. Finally, COC research on cognitive applications or services that are adaptive and perceptive are referred to as *Cognitive systems*. The summary of the basic lines of research on COC is summarized in Figure 4.

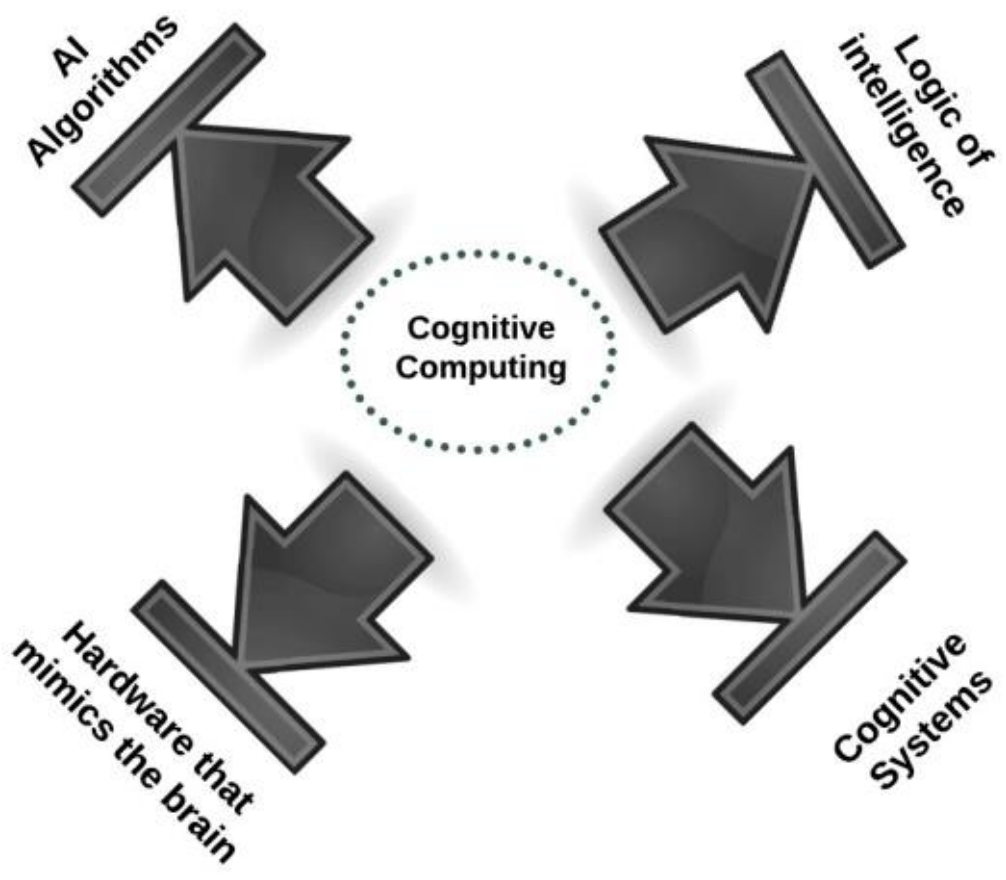

**Figure 4. Cognitive Computing Structure**

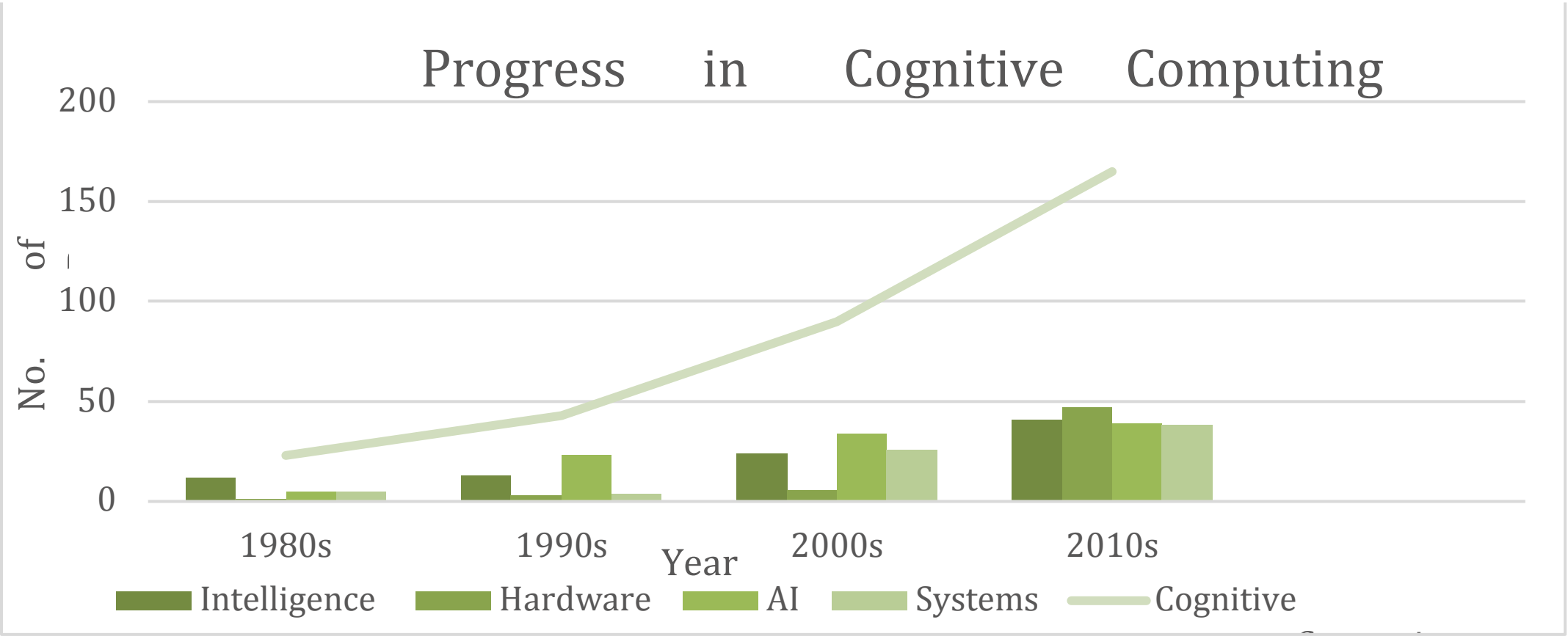

**Figure 5. Cognitive Computing Development from 1980 to 2020.**

Tracking the longitudinal development of research on COC as shown in figure 5, we notice a notable increase in the last two decades especially in the last decade. Research in the 1980's was mainly conceptual with a slight research in AI and Systems as a result of neural network emergence and Decision support systems. In the 1990's, research on AI was salient as the emergence of many AI algorithms such as genetic algorithms and clustering techniques. The 2000's had a slight increase in the brain-like hardware which has been boosted in the 2010's where the COC has dedicated research in the four lines.

## Cognitive Computing in Information Systems Research

One of the prominent findings from the analysis described in the previous section is the notable absence of IS related research on COC. Limited IS journals and conferences have referenced COC in their research. The IS basket of eight journals have barely mentioned COC as an emerging area of research with little focus on its conceptualization or how related research may be conducted (Watson 2017) (Sirinanda et al. 2017). While technology companies such as Google, IBM, and Microsoft are building high-tech cognitive systems for many organizations, IS research still considers COC as a future area of research!

As an evolving paradigm, it is crucial for Design Science Research (DSR) to build *a phenomenological epistemology* about COC. First, IS should build a knowledge base of the laws and the technological rules based on the current grounded knowledge about COC followed by generating and testing more specific "propositions". To build such epistemology, researchers should observe existing cognitive systems. *Reflection* is a type of thinking that aims for learning from experience (Cottrell 2003). Objectives in the *Reflection* thinking are set and action points that are saved into learning logs. Concrete experience is formed through active experimentation where reflective observations and abstract conceptualization are recorded (Cheng et al. 2017). *Reflection* will longitudinally investigate the experiences of individuals using COC.

Since there is little theorizing around COC, taking an *interpretative epistemological* approach (Smith 2004) could be a productive tactic to follow in designing behavioral research on COC. Moreover, Design science researchers could observe existing cognitive systems and reflect their findings. In other words, researchers could try to understand the outcomes of cognitive systems, how the outcomes are produced, and what influences COC use in organizational settings. Some important research questions for IS to pursue research on COC are shown in Table 2. The foreseeable effects on *Behavioral Science* and *Design Science* is discussed below:

### *Design Science*

since COC includes new forms of hardware, algorithms and logic. researchers should provide the guidelines of how to build cognitive artifacts such as models and instantiations under new hardware and more importantly, how to apply the new logic using traditional computers and the current AI algorithms such as deep learning algorithms to overcome current limitations of cognitive systems.

### *Behavioral Science*

basic activities in Behavioral science are: *Theorize and Justify*. *Behavioral science* aims to set theories, concepts, metrics which represent the theoretical base for IS research. This calls for IS to add relevant theories, methodologies, and logic such as *the neural theory of language* (Feldman 2008), *the computational theory of mind* (Horst 2011), and *the integrated theory of the mind* (Anderson et al. 2004). While the *justify* activity will only be validated when mature cognitive systems are fully developed and perform cognitive activities. *Justify* always reinforces theories and enhances generalizations. *Justification* in this case won't be only dependent on the social and technical implications of a theory, but also on the neural and psychological ones. The empirical research should be conducted on the COC artifacts where the settings should serve to verify the theory of interest.

| Future Research Questions | Description |
|---|---|
| How could COC contribute to IS behavioral science? | Behavioral science aims to set theories, concepts, metrics. Based on the conceptual representation of intelligence what will be the theories and metrics that characterize cognitive systems. What kernel and evaluation theories that could emerge to fit the COC structure? |
| How could COC contribute to IS design science? | Designing cognitive systems that adopt the COC structure would significantly be different from traditional intelligent systems which raise many questions such as what are design approaches that best fit the COC nature? In addition, how could IS build cognitive systems using the new *Neuromorphic Circuits*? |
| How could COC structure redefine what are called cognitive systems? | Cognitive systems in the light of the COC must have special characteristics that might differ from previously generated cognitive systems? To answer this question, there should be agreement on what constitutes a cognitive system. |
| How could COC *epistemological knowledge* of IS? | *Epistemological knowledge* includes, but not limited to, the establishment of abstract design knowledge about cognitive systems to guide IS initiatives to achieve the desired change. ContextMechanism-Outcome (CMO) could be a potential methodology to build such knowledge (Bonell et al. 2016). |

**Table 2. Research Questions for IS in COC.**

## Discussion and Limitations

Cognitive Computing is a cutting-edge advancement in computing that aims to empower machines with human cognitive abilities. However, the question of what the structure of COC is and what exactly is COC still remains unanswered. Moreover, there is very few IS research in COC. To answer these questions, we performed a systematic literature review using an inductive thematic analysis approach. The results of the analysis revealed the diversification of research areas in COC that could be grouped into four basic lines: logic of intelligence, hardware that mimics the brain, AI algorithms, and cognitive systems. However, no particular line could be considered COC, but all four lines are integral parts of COC structure.
Our analysis indicates that there is ample literature in each line that makes it not only an essential but an integral segment in the structure of COC. In other words, COC is not a single line of research but four integrated lines of research that include rudimentary or logic of intelligence, hardware that mimics the brain, AI algorithms, and cognitive systems. How the different lines of research complement each other is a question that requires further investigation. In addition, how these lines of research integrate and incorporate the overarching COC research structure calls for additional analysis.

Despite the significant advantages of TA (specially the theoretical flexibility to analyzing qualitative data), TA has limited interpretive power if it is not based on a clear theoretical framework. In addition, TA inherits the subjective dimensions of qualitative analysis. So, interpretations of TA are inevitably subjective and reflect the researcher's values and positioning (Boyatzis 1998). Moreover, TA also depends on the researcher's assessment of what is considered relevant. For example, researchers might avoid some themes that might be important because they thought them to be irrelevant. Therefore, diverse analysis methods are needed to reinforce our findings.

Given the fact that COC systems differ from previously generated cognitive systems, following an *interpretative epistemological* approach (Smith 2004) will guide IS researchers to conduct research in COC. This *interpretative epistemology* is accumulated through observing existing COC-based cognitive systems and reflecting researchers' findings to build a knowledge base about COC systems characteristics. Additionally, observing how the effects of COC-based cognitive systems are influencing organizational settings. Table 2 provides important questions for IS to pursue research in COC.

## Conclusion and Future work

Cognitive Computing (COC) is an emerging paradigm that is changing not only software systems but also the hardware structure of traditional computers. However, there is no agreement on what constitutes COC? Is it a special hardware, new AI algorithms, or an approach to develop systems? Indistinct delineation of what is COC contributed to inferior IS research in COC. So, this research aimed to explore the structure and nature of COC following a *Phenomenological Reflexives* approach using thematic analysis (Wisse 2003). Moreover, the research to propose guidelines for IS to conduct thoughtful research in COC. The findings of analysis suggest that there are four distinct yet integrated research lines of COC related to the logic of intelligence, hardware that mimics the brain (brain-like hardware), AI algorithms, and cognitive systems. However, future research is needed to investigate how the four lines are integrated. Using different analysis methods other than thematic analysis such as *Topic Modeling* and *knowledge graphs* would help verify the validity of our findings. The paper contributes to IS body of knowledge by shedding light on the foreseeable research questions on COC. In addition, the structure developed out of the thematic analysis could be considered a design science artifact that needs to be further evaluated (Peffers et al. 2012).